\documentclass[nonacm]{acmart}
\usepackage{balance}

\AtBeginDocument{%
  \providecommand\BibTeX{{%
    \normalfont B\kern-0.5em{\scshape i\kern-0.25em b}\kern-0.8em\TeX}}}

\setcopyright{rightsretained}
\copyrightyear{2021}

\acmConference[arxiv]{arXiV}{Fall, 2021}{New York, NY}
\acmBooktitle{arXiV, Fall 2021, Ithaca, NY}

\usepackage[normalem]{ulem}
\useunder{\uline}{\ul}{}
\usepackage{booktabs}
\usepackage{subcaption}
\usepackage{threeparttable}
\usepackage{multicol}
\begin{document}
\title{The PUEVA Inventory: A Toolkit to Evaluate the Personality, Usability and Enjoyability of Voice Agents}

\author{Stacey Li}
\authornote{Both authors contributed equally to the paper}
\email{stacey.li@macaulay.cuny.edu}
\affiliation{%
  \institution{Hunter College and Cornell Tech}
  \streetaddress{2 W Loop Rd}
  \city{New York}
  \country{USA}}
  
\author{Sven Krome}
\authornotemark[1]
\email{svenkrome@gmail.com}

\author{Ilan Mandel}
\email{im334@cornell.edu}
\affiliation{%
  \institution{Cornell Tech}
  \streetaddress{2 W Loop Rd}
  \city{New York}
  \country{USA}}
  
\author{Marcel Walch}
\email{marcel.walch@alumni.uni-ulm.de}
\affiliation{%
  \institution{Ulm University}
  \city{Ulm}
  \country{Germany}}

\author{Wendy Ju}
\email{wendyju@cornell.edu}
\affiliation{%
  \institution{Cornell Tech}
  \streetaddress{2 W Loop Rd}
  \city{New York}
  \country{USA}}

\settopmatter{printacmref=false}

\renewcommand{\shortauthors}{Li and Krome, et al.}

\begin{abstract}
The proliferation of voice agents in consumer devices requires new tools for evaluating these systems beyond their technical functionality. This paper presents a toolkit for the evaluation of Voice User Interfaces (VUIs) with the intention of measuring the crucial factors of subjective enjoyment in the user experience. The PUEVA toolkit was constructed using a meta-analysis of existing literature and a within subjects lab study. The resulting questionnaire contains 35 items that represent 12 scales in three categories: (1) \textit{Personality} (2) \textit{Usability} and (3) \textit{Enjoyability}. The PUEVA Toolkit moves us towards the capacity to evaluate and compare subjective, joyful experiences in between-subject as well as within-subject research designs.
\end{abstract}





 \maketitle

\section{Introduction}
Voice agents which interact via natural speech input and respond using a synthesised voice are commonly found in mobile phones and home speakers. Recently, voice agents have made significant inroads into homes, offices and automobiles \cite{corbett2016can}. As natural language generation (NLG) has improved, voice agents have been imparted with additional personality traits expanding interaction beyond the mechanized style of earlier voice agents\cite{chen2010behavior}. 

Given the number of new applications that use voice agents as an interface, we felt that the community needed a common toolkit for evaluating the perceived enjoyment and user experience of interactions with voice agents, one that could be used across different voice agent applications. Few studies exist on what people do to overcome VUI problems they encounter. In this paper, we present the design of an inventory questionnaire intended to help designers and researchers working on novel voice user agents to evaluate their designs. Moreover, the importance of specific aspects of existing scales may have changed. Our new scale respects prior academic work by building on previous scales, and makes it easier to compare systems evaluated with older scales with future systems evaluated using our scale. We drew on existing literature and existing scales to test a super-set of current scales, and used principal component analysis to reduce these to a more practicable toolkit of questions.

\section{Related work}
\subsection{Voice User Interfaces}
While there are pragmatic reasons for the increase of voice interfaces on a variety of consumer devices, the ubiquity of voice user agents is often related to the desire to make for more ``natural'' user experiences  \cite{castelluccio2019conversational}. Interviews and surveys with people comparing qualities of good human conversation with good human-agent conversation indicate that people evaluate human conversations around relationship building and maintenance, but have more transactional standards around interactions with voice agents. \cite{clark2019makes} 

That said, a substantial body of prior research on voice user interfaces indicate that factors beyond mere task execution color the experience and enjoyment people have interacting with a voice agent \cite{nass2001does, bonfert2018if}. 
Interactive technologies are capable of triggering our natural social responses to speech \cite{nass2005wired} which necessitate tools that allow for evaluation along dimensions of both usability and personality.

While numerous studies exist evaluating the usability and user experience of a variety of VUIs \cite{abdolrahmani2018siri, pyae2018investigating, lopez2017alexa, ehrenbrink2017google}, often for different populations or domains \cite{lovato2015siri,corbett2016can, mennicken2018challenges, sayago2019voice, pal2019user, kowalski2019older,augstein2019weldvui}, these evaluations are often difficult to compare or cross apply because they use different measures and scales to evaluate the interactions. Developing and leveraging a more standardized questionnaire that captures both usability and personality would allow for cross-context VUI prototype evaluation, comparison and the development of design best practices for voice agents. 

\subsection{Evaluation as Part of Voice User Interface Design}

Many of the HCI studies performed on existing consumer voice interfaces are executed by academics after a product is shipping, but, from the perspective of this paper's authors, the most useful phase of the product life cycle in which to perform evaluation is when interactions are being authored. This point-of-view argues for a lightweight and flexible evaluation tool that allows people to compare different platforms \textit{(Should we build use Cortana or Google Voice for this automotive application?)}, different use cases \textit{(Does this voice application work as well in English as it does in Japanese?)}, different voices \textit{(Should we use a male or a female voice for this?)}\cite{cambre2020choice} or different interaction designs \textit{(Should we make our voice agent more polite, or more assertive?)} around common subjective factors. 

At each stage of prototyping, VUIs must be evaluated while accounting for the conversational interaction paradigm.\cite {murad2019revolution}   
 We believe VUIs to be more like mobile apps, which tend to want to be studied in the context where they are used, or webpages, which are easily deployed at scale. Therefore, we believe a questionnaire-based evaluation tool, wherein the users are asked to self-evaluate the qualitative aspects of the voice agent following an actual interaction with the system, are more appropriate than heuristic-based tools for making key design decisions.

\subsection{Questionnaires for Usability, UX and Interaction}
While novel interfaces or systems sometimes require bespoke evaluation instruments intended to ensure that the designed system fulfills its intended purpose, early HCI researchers found that it was useful to develop standardized evaluation instruments to measure different software with a common yardstick.   
Early instruments such as the Questionnaire for User Interface Satisfaction \cite{chin1988Development}, IBM's Afer-Scenario Questionnaire (ASQ) \cite{lewis1995ibm}, and the Post-Study System Usability Questionnaire (PSSUQ) \cite{lewis1992psychometric} all had strong internal consistency, but they often focused on user's self assessment of task completion and satisfaction, and often were not sensitive enough to pick up on the effects of various design changes \cite{hornbaek2006current}. Later questionnaires broadened to consider adaptation to other contexts (for example, educational technology \cite{harrati2016exploring}, mobile applications \cite{ryu2006reliability} and medical instruments \cite{liljegren2006usability}).

One of the major departures from the early HCI questionnaires was the shift in focus from usability and task-oriented measures to user experience and other more aesthetic measures \cite{tractinsky1997aesthetics, hassenzahl2003attrakdiff, laugwitz2008construction, schrepp2017design}. 
In contrast, researchers in the communication and voice-based research space often borrow standard psychometric and relational scales from psychology and sociology \cite{reeves1996media, nass2005wired}.


In light of the variance of possible perspectives on appropriate measures for speech interfaces, then, we cast a wide net in sourcing and testing multiple standards scales upon which to base our toolkit to evaluate the personality, usability and enjoyability of voice agents.

\section{Questionnaire Development}
\subsection{Literature Meta-Review}
A review of literature was conducted to establish an understanding of the most relevant scales and measures of VUIs. The review not only included canonical VUIs literature but also relevant literature from social-robot interaction and embodies voice agents. The goal was to gain an overview of research on VUIs and related domains and to identify the most important scales and measures in this domain. Based on combinations of key-words (voice interaction, social robots, affective computing, conversational interfaces, speech interfaces etc.) and a number of citations from research databases, relevant papers were collected and reviewed focusing on the goodness criteria for VUIs. Additionally a systematic review of the conference proceedings from the last five years including the conferences CHI, CSCW, AutoUI, DIS, MobileHCI and IUI. The goal of this material was to identify what evaluation methods have been applied in the most recent studies on VUIs.

\subsubsection{Paper Clusters}
We cluster the literature into five categories discussed briefly below.

\paragraph{Voice User Interface In Everyday Life}
Qualitative interviews and ethnographic work looking at in-situ use of VUIs is informative for understanding how these systems are being put in use today. Interviews have shown significant gaps between user expectations and device capabilities \cite{luger2016like}. Months long evaluations of VUIs placed in the home have studied how the device may be made accountable and embedded in multi-party interactions \cite{porcheron2018voice}. The social experience of an interaction with a VUI shows wide variation. Analysis of user reviews of the Amazon Echo and its VUI Alexa show significant divergence in user's personification of the device and the names or pronouns they choose to refer to it \cite{purington2017alexa}. Surveys of existing users of voice agents elucidates challenges with the usability of these systems that may present areas for technical improvement \cite{pyae2018investigating}. While voice user agents have existed for decades, more advanced and mainstream VUIs such as Siri, Alexa and Google Home attempt to be significantly more interactive, requiring new forms of interaction design.

\paragraph{Conversational Skills and Personality}
A large body of work in conversational analysis is being leveraged for the development of Conversational UX as a distinct design space \cite{moore2017conversational}. Conversational flow and turn taking has been explored by researchers interested in VUIs as well has in Human Robot Interaction (HRI). When interacting with limited systems, people will often modulate their own speech in response \cite{pelikan2016nao, porcheron2018voice}. The capacity for VUIs to interact in a way people find natural has implications for the devices utility to users.

\paragraph{Usability}
Comparisons of the four most popular voice assistants show that on metrics of both naturalness of interaction and usability, these systems still leave a lot to be desired \cite{lopez2017alexa}. Measures of system performance are widely varied \cite{mctear2016evaluating} and can be separated into overall system evaluations and component evaluations. An example of an overall system evaluation is the PARAdigm for DIalogue Evaluation System (PARADISE), a weighted combination of metrics which capture task success, dialog efficiency, dialog quality, and user satisfaction \cite{walker1998evaluating}. Component level evaluation is a collection of numerical metrics for the functionality of VUI technology such as word error rate for the evaluation of Automatic Speech Recognition. 

\paragraph{Joy and Engagement}
Multiple scales can be considered when evaluating a participants’ perception of VUIs. For example, a participant’s engagement with an autonomous system can be measured by ability to recall details of a story, the perception of a robot, and any self-reported learning \cite{szafir2012pay}. Learnability and discoverability can be seen as a medium in which engagement is fostered. The learnability of a VUI is most challenging during a user’s initial encounter with it \cite{corbett2016can}. Researchers \cite{qvarfordt2003role} found that people prefer more tool-like interactions but if they experienced a good alternative, would begin to prefer a more human-like interaction. The quality of engagement can be an indicator of joy that a user experiences. 

\paragraph{Long Term Interaction and Relationships}
VUIs can be understood as a category of relational agents, defined by Bickmore and Picard \cite{bickmore2005establishing} as ``computational artifacts designed to establish and maintain long-term social-emotional relationships with their users''. VUIs reside in users devices and their homes and have the potential to maintain interactions for extended periods of time.  Attempts have been made to explicitly design for long term companionship in interactive systems \cite{benyon2007introducing}. A VUI in a shared common space acting as a relational agent can affect the overall group dynamics of the people who interact with it \cite{lee2019hey}.

\subsubsection{Constructs}
From the literature review we identified the following constructs as critical for the
evaluation of the experience of VUIs:
\begin{itemize}
    \item Positive and negative emotions triggered by the system 
    \item User acceptance of the system
    \item The relationship between the user and the system
    \item The attitudes towards the system
    \item The personality and character conveyed by the system
    \item Companionship features of the system
    \item The perceived intelligence of the system
\end{itemize}
With an understanding of the relevant constructs for interaction with voice agents, we looked at the validated questionnaires used in our Meta-analysis.

\subsection{Questionnaire Super-Set}
We collected a set of validated measurement instruments such as scales and diagnostic tools that address various aspects of VUI evaluations. The collection includes VUI evaluation tools as well as scales from related domains such as psychology and sociology as well as industry standards. We combined the following into our super-set: 
\begin{itemize}
    \item SASSI - Subject Assessment of Speech Interfaces \cite{hone2000towards}
    \item UTAUT Scale - Unified Theory of Acceptance of Technology \cite{heerink2010assessing}
    \item WAI - Working Alliance Inventory \cite{horvath1989development}
    \item RAS - Robot Anxiety Scale and NARS - Negative Attitudes towards Robots Scale \cite{nomura2008prediction}
    \item NEO-Five-Factor Inventory \cite{costa1995persons}
    \item CSAP - Companionship Scale for Artificial Pets \cite{luh2015development}
    \item Godspeed Scale \cite{bartneck2009measurement}
\end{itemize}

\subsection{Dimension Reduction}
 After collecting all items from the standardized questionnaires above, we had a table consisting of 235 items that describe 50 scales. We then removed redundant items along with items and scales that could not be adapted to VUI scenarios. Furthermore, we removed or rearranged the scale of items with the objective that all remaining items are rated by a 5-point Likert scale. 

The resulting 103-item questionnaire super-set is still too long to be useful in most evaluation scenarios. Hence, we administer the super-questionnaire to participants following interactions with VUIs, and used Principal Component Analysis (PCA) to determine the component loadings and reduce the number of questions.

\subsubsection{Methods} Participants (\textit{N}=13) recruited through convenience sampling were individually brought into a room containing a Google Home, an Alexa Dot, and an Apple Home Pod. The researcher presented them with the wake word for each device and a Demo Request such as ``Hey Siri, how are you.'' Participants were asked to interact with the three systems for at least 10 minutes. A guide of possible requests and tasks was provided but the participants were instructed to be free and spontaneous in their interaction. The researcher left but monitored the interactions remotely. After 10-15 minutes (depending on the participants activity of interaction with the
system) the researcher entered the room and provided a digital version of the 103-item questionnaire with the order randomized. Participants were asked to fill out the survey and were free to continue interacting with the device as they did so.

\subsubsection{Principal Component Analysis}
\begin{figure}
    \centering
    \includegraphics[width=.9\linewidth]{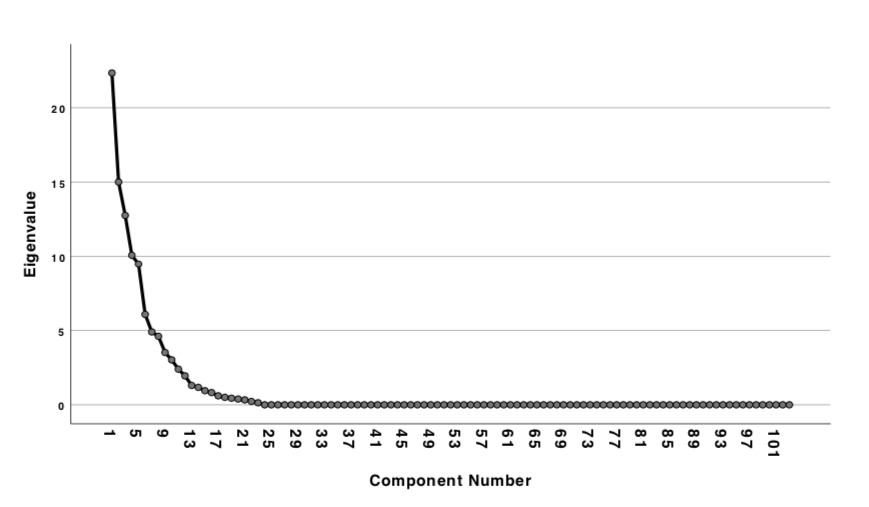}
    \caption{Scree Plot}
    \label{fig:scree}
\end{figure}

We assume that many of our factors are correlated and performed principal component analysis with  direct oblim rotation. As illustrated in the scree plot in Figure \ref{fig:scree}, this results in 14 components with eigenvectors greater than one that explain more than 95\% of the total variance in our data. 

We conducted component analysis iteratively, removing components with loadings below 0.6 before conducting the PCA again.

\subsection{Synthesis}
\begin{table*}[]
\caption{The PUEVA Questionnaire Items and Scales}
\subcaption*{Personality}
\begin{tabular}{@{}p{.27\textwidth}p{.25\textwidth}p{.23\textwidth}p{.19\textwidth}@{}}
\toprule
  \textbf{Conversational Skills Scale} &
  \textbf{Playfulness Scale} &
  \textbf{Happiness Scale} &
  \textbf{Friendliness Scale} \\ 
  
  \emph{The degree to which the system can hold a coherent conversation}
  &
  \emph{The degree to which the system conveys lighthearted personality traits} &
  \emph{The degree to which the system conveys is own positive or negative emotional affect.} &
  \emph{The degree to which the system is perceived as affable.}  \\\midrule
The SYSTEM is upright and honest. &
  *The SYSTEM is responsible. &
  *The SYSTEM seems depressed &
  The SYSTEM is kind. \\
The SYSTEM seems energetic.&
  The SYSTEM is foolish. &
  The SYSTEM seems cheerful &
  The SYSTEM is pleasant. \\
The SYSTEM has its own opinions. &
The SYSTEM seems self-conscious &
   &
   \\
I consider the SYSTEM a pleasant conversational partner. &
  *The SYSTEM seems assertive. &
   &
   \\ \bottomrule
   \end{tabular}
\newline
\subcaption*{Usability}
\begin{tabular}{@{}p{.27\textwidth}p{.25\textwidth}p{.23\textwidth}p{.19\textwidth}@{}}
\toprule
  \textbf{System Performance Scale} &
  \textbf{Speed Scale} &
  \textbf{Understanding Scale} &
  \textbf{System-State Scale} \\ 
  
  \emph{The degree to which the system behaves as expected and performs effectively} &
  \emph{The efficiency of the system} &
  \emph{The capacity for the system to interpret and adapt to interaction.} & \emph{The degree to which the system can perform consistently.} \\ \midrule
*The SYSTEM didn’t always do what I wanted. &
  I found the interaction with the SYSTEM to be fast. &
  The SYSTEM understands difficult conversation topics. &
  I was able to recover easily from errors. \\
*The SYSTEM didn’t always do what I expected &
  The SYSTEM always gives me an immediate response. &
  The SYSTEM adapts well to new situations. &
  The SYSTEM is well-organized. \\
A high level of concentration is required when using the SYSTEM &
   &
   &
   \\
I think I can use the SYSTEM without any help &
   &
   &
   \\ \bottomrule
      \end{tabular}
\newline
\subcaption*{Enjoyability}
\begin{tabular}{@{}p{.32\textwidth}p{.32\textwidth}p{.32\textwidth}@{}}
\toprule
  \textbf{Pleasure Scale} &
  \textbf{Companionship Scale} &
  \textbf{Usefulness Scale}\\ 
  
  \emph{The degree to which users enjoy using the system} &
  \emph{The degree to which the system can be treated like a friend.} &
  \emph{The perceived utility of the system} \\ \midrule
  
The SYSTEM makes me happy when I talk with it. &
  Whether I am happy or sad, I want to share it with the SYSTEM. &
    I believe the time spent with the SYSTEM is not spent efficiently. 
   \\
The SYSTEM provides me with pleasurable activity. &
  The SYSTEM is like a friend who can keep me from being lonely. &
  I think the SYSTEM will help me when I need help. 
   \\
I enjoy the SYSTEM talking to me. &
   &
I found the SYSTEM to be useful. 
   \\
I find the SYSTEM fascinating. &
   &
  I am confident in the SYSTEM's ability to help me.
   \\
I like the SYSTEM. &
   &
  The SYSTEM is intelligent. 
   \\
The SYSTEM can make life more interesting &
   &
   \\ \bottomrule
\end{tabular}

\begin{tablenotes}
  \centering
  \item *Items with reversed scales
\end{tablenotes}
\label{pueva}
\end{table*}

Using the results from the principal component analysis as inspiration rather than a decisive criteria, we developed the finalized questionnaire. To do so, we ran several iterations keeping only items with a larger loading than 0.6. Analysing and comparing the components, we constructed the final questionnaire. A few ambiguously worded questions were removed or substituted for highly-correlated items from the structure matrix. 

The resulting questionnaire comprises 35 five-point Likert questions grouped in 11 scales as listed in Table 1. Each of the scales consists between 2 to 5 items. We assigned the scales into three categories depending of the inquiry target. Personality, for scales that assess the characteristic of the system. Usability, for scales that measure the functionality of the interactions with the system and Enjoyment, consisting of the scales that measure the experiential impact of the system.

\section{Discussion}
The PUEVA Inventory standardizes the process by which VUIs can be developed on the basis of personality, usability, and enjoyability. While the open-ended survey and interview of people's interactions with voice agents contained many interesting anecdotes and heartfelt feelings, they also took a lot of time to process and map to themes. We believe that the inventory design covers many of the key issues that influence judgement of voice agents. Moreover, the inventory has been tested to be understandable and easy to use. Whereas the 103-item questionnaire took up to 15 minutes to complete, the 35-item PUEVA inventory should be deployable within 5 minutes.  

\subsection{Comparison of PUEVA to other scales} 

One of the motivations for the PUEVA scale is that many existing assessment methods for Voice User Interfaces \cite{hone2000towards} are oriented toward basic voice input operations rather than the variety of more fully fledged interactive agents increasingly at use. Leveraging these scales as a basis but including subjective measures from fields such as Human Robot Interaction (HRI) and social psychology helps us to maintain the face and construct validity of our questionnaire while expanding their usage toward modern VUI applications. As a whole, there are no existing scales like PUEVA, which seeks to evaluate how users perceive voice agents, whose interactions are becoming increasingly human-like. 

How similar is PUEVA to any of its constituent scales? This is an important question, even if we feel the PUEVA scale adds important dimensions to the overall analysis of interactive agents, as high similarity might indicate that comparisons with prior research using prior research is viable. Of all included inventories, the SASSI questionnaire seems predictably to be the most similar in terms of the structure of the scales, despite the fact that only 6 SASSI items made it into the PUEVA inventory. Almost all (i.e. 5 of the 6) SASSI items, were loading on our usability scales, particularly the System Performance Scale. The only exception is the item "The SYSTEM is pleasant" which had a high factor load on our Friendliness component. Surprisingly, none of the Annoyance items of the SASSI inventory was loading significantly on our 11 identified components. In other words, despite some overlap in questionnaire items, the PUEVA inventory measures  very different contructs than SASSI overall. 

On the whole, the PUEVA inventory emphasizes personality- and relationship-related constructs over usability- and performance-related constructs. We believe that the shift in focus was impacted by two phenomena. First and foremost, the commercial VUI systems that we used in conducting the studies that informed the PUEVA inventory questions and weights are much more mature than the system from the time of SASSI's development in 1999/2000. Secondly, the basis for constructing PUEVA consisted of a much more varied sources of questions items, including personality and relationship oriented inventories. Finally, from a methodological perspective, we did not give our participants a definitive task to perform but motivated them to "toy around" with the system at will. This focus seemed to orient the evaluation towards more positive, emotional or relation-able constructs of the source inventories, as our participants were able to explore the pleasurable and communicative qualities of the systems. Looking beyond the usability of the VUI systems, we believe that the PUEVA questionnaire helps to assess the actual relational quality of the interacting in natural language with an artificial system by covering the key aspects discussed in literature: personality, usability and enjoyability. 

\subsection{Other possibilities for extension of PUEVA} 

As we mentioned, we know the PUEVA inventory does not cover all the factors that we discovered that people care about in our interviews:  Visuals and embodiment, Users' self-reflection on how they communicate with the system, Structure the day, Pleasure of venting/aggression. We believe visuals and embodiment to be particularly important factors, which are currently not addressed by our inventory. For example, if, in a car, a person asks for nearby restaurants, it is beneficial if the agent can show a listing of restaurants and relative distances, or locations on a map; these are all better than reading out each listing aloud. 

While PUEVA was developed for immediate contextual feedback on design, we feel there are opportunities to extend the scales to address longitudinal use. The playfulness of a brief interaction may become annoying with extended use. To that end, we acknowledge the need for additional items and/or additional scales to capture factors pertaining to longer-term interaction and usage.



\section{Conclusion}
 The main contribution of this work is a research and assessment instrument, the PUEVA Inventory, designed to aid in evaluating pleasurable experiences of voice user interface by novice users based on short-term use. This inventory was built on an exhaustive list of scales and measures that can be used to evaluate relevant HCI systems. We believe this instrument to be particularly useful for current-day voice interactions, which have greater focus on personality and enjoyment than the more basic task-oriented voice interfaces of old. With an increasing number of VUIs emerging onto the market, we hope researchers will use the PUEVA Inventory to make friendlier, more approachable and more enjoyable interfaces for us all to use.

\begin{acks}
We'd like to acknowledge Mitsubishi Motors, especially Zacharias
Cieslinski, for their support in this research.
\end{acks}

\balance
\bibliographystyle{ACM-Reference-Format}
\bibliography{Pueva}


\end{document}